\documentclass{vgtc}                          % final (conference style)
\ifpdf%                                % if we use pdflatex
  \pdfoutput=1\relax                   % create PDFs from pdfLaTeX
  \usepackage{graphicx}                % allow us to embed graphics files
  \DeclareGraphicsExtensions{.pdf,.png,.jpg,.jpeg} % for pdflatex we expect .pdf, .png, or .jpg files
\else%                                 % else we use pure latex
  \usepackage{graphicx}                % allow us to embed graphics files
  \DeclareGraphicsExtensions{.eps}     % for pure latex we expect eps files
\fi%

\graphicspath{{figures/}{pictures/}{images/}{./}} % where to search for the images

\usepackage{microtype}                 % use micro-typography (slightly more compact, better to read)
\PassOptionsToPackage{warn}{textcomp}  % to address font issues with \textrightarrow
\usepackage{textcomp}                  % use better special symbols
\usepackage{mathptmx}                  % use matching math font
\usepackage{times}                     % we use Times as the main font
\usepackage{cite}                      % needed to automatically sort the references
\usepackage{tabu}                      % only used for the table example
\usepackage{booktabs}                  % only used for the table example
\usepackage{url}

\onlineid{0}

\vgtccategory{Research}

\vgtcinsertpkg

\usepackage{eso-pic,xcolor}
\usepackage{lipsum} % dummy text

\usepackage{tikz}

\title{Extended Reality and Internet of Things for \\Hyper-Connected Metaverse Environments}

\author{Jie Guan \thanks{e-mail: jie.guan@ocadu.ca} %
\and Jay Irizawa  \thanks{e-mail:jirizawa@ocadu.ca} %
\and Alexis Morris\thanks{e-mail:amorris@ocadu.ca}}
\affiliation{\scriptsize Adaptive Context Environments (ACE) Lab \\ OCAD University}                                                                               

\abstract{
The Metaverse encompasses technologies related to the internet, virtual and augmented reality, and other domains toward smart interfaces that are hyper-connected, immersive, and engaging. However, Metaverse applications face inherent disconnects between virtual and physical components and interfaces. This work explores how an Extended Metaverse framework can be used to increase the seamless integration of interoperable agents between virtual and physical environments. It contributes an early theory and practice toward the synthesis of virtual and physical smart environments anticipating future designs and their potential for connected experiences.

} % end of abstract

\keywords{Metaverse, mixed reality, internet-of-things, agents.}

\CCScatlist{
  \CCScat{Human-centered computing -- }{Mixed / augmented reality}{}{};
  \CCScat{Human-centered computing -- }{Virtual reality}{}{};
  \CCScat{Human-centered computing -- }{Ambient intelligence}{}{};
}

\usepackage{tikz}
\usepackage{lipsum}

\begin{document}
\begin{titlepage}

     \vspace{1cm}
        Full Citation: J. Guan, J. Irizawa and A. Morris, "Extended Reality and Internet of Things for Hyper-Connected Metaverse Environments," 2022 IEEE Conference on Virtual Reality and 3D User Interfaces Abstracts and Workshops (VRW), Christchurch, New Zealand, 2022, pp. 163-168, DOI: 10.1109/VRW55335.2022.00043.

       \vspace*{1cm}

       \copyright2022 IEEE. Personal use of this material is permitted.  Permission from IEEE must be obtained for all other uses, in any current or future media, including reprinting/republishing this material for advertising or promotional purposes, creating new collective works, for resale or redistribution to servers or lists, or reuse of any copyrighted component of this work in other works.

       \vspace{1.5cm}
  
\end{titlepage}

\firstsection{Introduction}

\maketitle

%% \section{Introduction} %for journal use above \firstsection{..} instead
The concept of the Metaverse is increasingly becoming adopted into everyday society with ubiquitous applications from common living and working scenarios, to customized forms of social activities enhancing a connectedness of experiences.  %engagements of social activities, entertainment, and other forms of activities%, wherein humans are now gradually engaged more and more in the Metaverse. 
The Metaverse is challenging to define, however, across multiple definitions it is often overlapped with concepts of an environment constructed by multiple and discrete virtual worlds, defined as a fully immersive, three-dimensional digital environment that reflects the totality of shared online space\cite{Dionisio20133D}. Lee et al.\cite{Lee2021All} consider the Metaverse as a virtual environment constructed by the Internet, Web technologies, and Extended Reality (XR) toward hybrid physical and virtual space.
%In this sense, XR is a concept allows for visualizing the Metaverse content in both Immersive virtual and hybrid environments. 
In this sense, XR is an expanded field of fluid space enabling the visualization of Metaverse content in both immersive virtual and hybrid environments. It is an umbrella term for  Virtual Reality (VR), Augmented Reality (AR), and Mixed Reality (MR) of the virtuality continuum concept \cite{Milgram1994taxonomy} which is a mixture of the presenting of objects from-real-to-virtual displays. XR is also a term that refers to the combination of physical and virtual (computer-generated graphic) environments that humans can interact. In particular, "X" represents the spatial computing technologies connecting virtual and physical space\cite{Greenwold2016Spatial}. 

The Metaverse is a growing concept at this time, yet is in a state of rapid development approaching mainstream acceptance in commercial and consumer applications. However, it is often envisioned with the premise that a more mature ubiquitous Metaverse will provide many benefits for humans to connect -- in various spaces and with each other, creating synchronous social engagements overcoming physical distance and potential limitations that expend time and energy. In the most recent developments from fields of academia and industry, there is evidence to support this trend toward Metaverse research and integration. For example, in 2021, Facebook announced that they were becoming a Metaverse company\cite{facebookMARKINTHEMETAVERSE} and rebranded itself to Meta Platforms Inc. Other major industry influencers, Roblox and ZEPETO are also representative of social platforms that have enormous users across the world and have been moving toward the adaptation of Metaverse capabilities \cite{han2021analysis}. Likewise, Nvidia has further developed its Omniverse software\cite{NVIDIAOMNIVERSE} for creators to simulate the constraints and details of the physical space using a digital twin 3D connected simulation. Examples like these large-scale industry developments show how the Metaverse has not only become one of the most influential trends in the marketplace; various industries are adapting their services to provide the foundations of networked platforms for an expanded collaborative Metaverse space.

Similarly, with the mainstream adoption of virtual reality hardware and software, users can now enter diverse Metaverse platforms of virtual spaces with stand-alone computers, mobile devices, and head-mounted VR display devices. Such devices transport users into completely virtual spaces, but challenges arise with their use. For example, with the advent of multiple virtual shared spaces, there comes a disconnect -- there is significant development of virtual content within the Metaverse, but much less focus on the connections of this content within the physical space (i.e., the user's actual environment). However, regardless of how we engage with the virtual through conventional virtual reality displays, humans still rely on physical environments for basic needs to function and survive. As a consequence, if the Metaverse does not take into consideration the design to connect and maintain coherency with the physical-virtual relation as a hybrid construct, the Metaverse will not be optimized to fully support human activities in any of these environments, co-opting a sustained and mutually exclusive distinction between physical and virtual worlds.

\subsection{Metaverse Disconnect: A Research Opportunity}
\begin{figure}[tb]
 \centering % avoid the use of \begin{center}...\end{center} and use \centering instead (more compact)
 \includegraphics[width=\columnwidth]{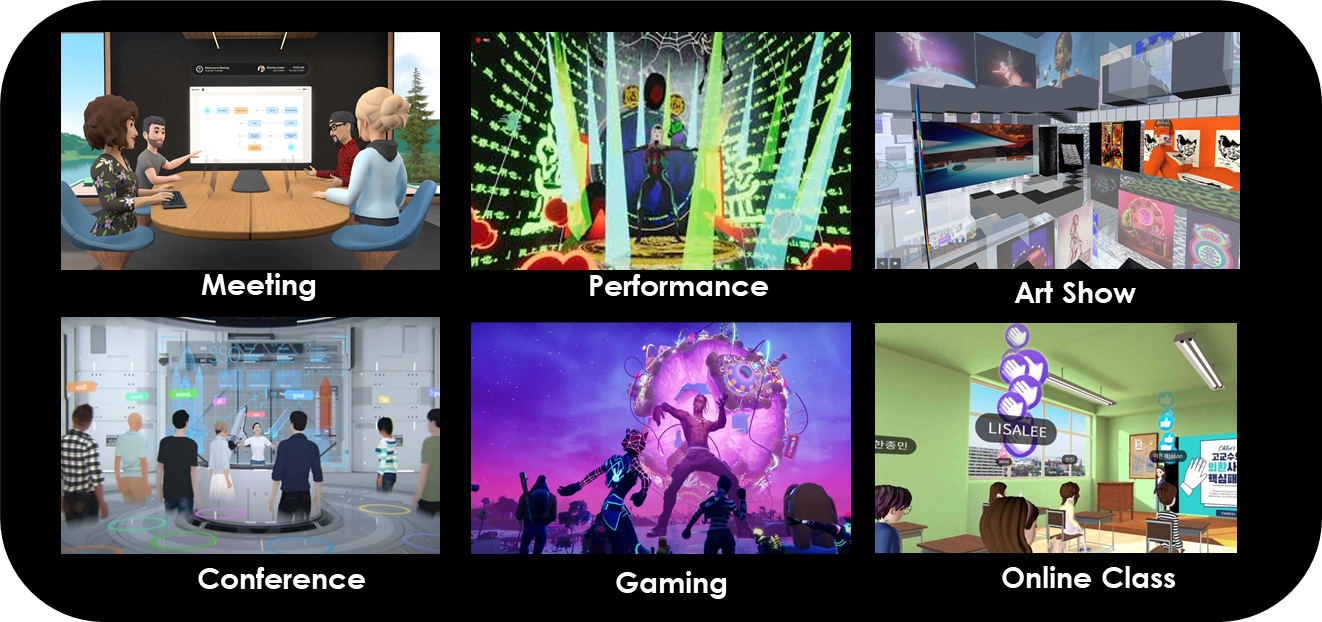}
 \caption{Examples of current activities in today's metaverse applications (e.g., Horizon Worlds\cite{horizonworlds}, Decentraland\cite{decentraland}, Cryptovoxels\cite{cryptovoxels}, Spatial\cite{spatial}, Fortnite\cite{Fortniteepicgames}, HanaBank\cite{HanaBankmin-kyung_2021}).}
 
 %\footnote{Horizon Worlds: https://www.oculus.com/horizon-worlds/;\\ Decentraland:https://decentraland.org/;\\ Cryptovoxels: https://www.cryptovoxels.com/;\\ Spatial: https://spatial.io/;\\ Fortnite: https://www.epicgames.com/fortnite/;\\ Hana Bank: http://www.koreaherald.com/view.php?ud=20210806000630.}

%  \caption{Examples of current activities in today's metaverse applications, including Horizon Worlds\footnote{https://www.oculus.com/horizon-worlds/}, Decentraland\footnote{https://decentraland.org/}, Cryptovoxels\footnote{https://www.cryptovoxels.com/}, Spatial\footnote{https://spatial.io/}, Fortnite\footnote{https://www.epicgames.com/fortnite/en-US/home?sessionInvalidated=true}, Hana Bank\footnote{http://www.koreaherald.com/view.php?ud=20210806000630}.}
 \label{CurrentActivitiesinMetaverse}
\end{figure}

Today, Metaverse platforms in web, mobile, and VR devices have been used for various social activities, for meetings, sports and entertainment, exhibitions, conferences, gaming, tourism, and online education (see \autoref{CurrentActivitiesinMetaverse}). However, the current Metaverse platforms are limited in their capacity to support meaningful immersion and duration for people to stay connected. Without a strong and continuous sense of connection, the reliability of a coherent Metaverse will remain a elusive, suggesting a need to consider the design for the interdependent relationship of the virtual content and world/s, to the physical world/s people inhabit.

\begin{figure}[tb]
 \centering % avoid the use of \begin{center}...\end{center} and use \centering instead (more compact)
 \includegraphics[width=\columnwidth]{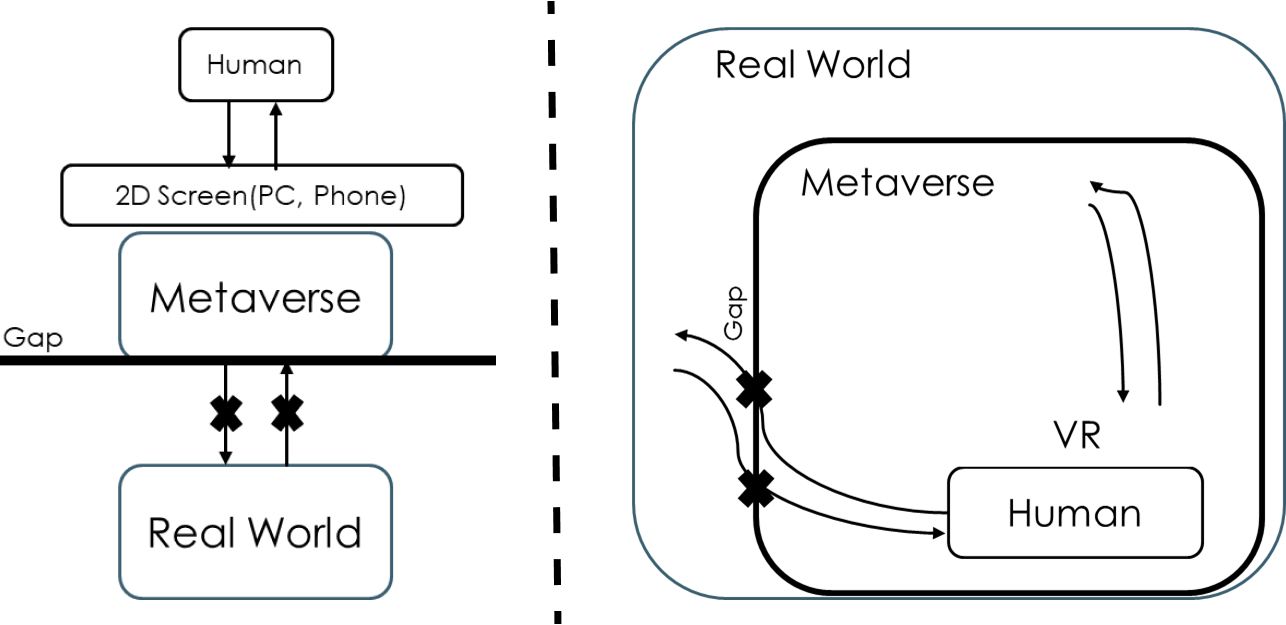}
 \caption{Current Metaverse interaction and the Metaverse disconnect problem (adapted from early explorations on VR by Rekimoto \cite{Rekimoto1995world}). There is a need to more strongly link the Metaverse to the real-world.}
 \label{CurrentMetaverseInteraction}
\end{figure}

This work proposes the following question: how can the gap between the Metaverse and the physical world be minimized to increase the dynamic engagement for smart environments with Mixed Reality and Internet-of-Things? Augmented Interaction from human-computer interaction (HCI) research \cite{Rekimoto1995world} highlights this issue, and the need for the right computing paradigm to support virtual and physical areas of interaction. As shown in \autoref{CurrentMetaverseInteraction}, when users apply two-dimensional screen or Virtual Reality devices to enter Metaverse applications or environments, they face a naturally occuring gap with the real world, as a result of the platform being used. Often, this means that, for virtual environment interactions, they can not access and manipulate the physical information found in real-world environments surrounding them. Thus the Metaverse disconnect problem is a significant hurdle for the field to overcome as it evolves toward mainstream adaptation.

This paper explores the design of an Extended Metaverse Framework to connect the physical and virtual space using Mixed Reality (MR) and the Internet-of-Things (IoT). The aim is to enhance the interconnection of the Hyper-connected smart environment. In doing so, this work describes how three-dimensional interface (virtual content) designs in MR can allow users to interact with the physical elements in a smart-environment through the IoT. 
This interaction allows users to switch between an immersive Metaverse environment and the MR (users can see/sense the environment they are physically inhabiting) space. On the other hand, the user could control the physical objects such as lighting fixtures by manipulating the immersive virtual interface. %The discussion and future work present the possibility of using the L-system as the content creation method to generate virtual content to prevent the overloading of Metaverse by feeding physical information and rules with affect virtual.

This paper is presented as follows: Section 2 discusses the background related to the Metaverse and the themes involved in mixed reality and IoT. Section 3 presents the Extended Metaverse Framework and early designs of Metaverse application concepts as a proof-of-concept. Section 4 provides a discussion of these techniques, and Section 5 summarizes the paper.

\section{Metaverse: Theory and Background}
 
% Literally, Metaverse is a portmanteau that combines with the prefix ``meta", which means ``beyond," and the suffix ``verse," which is the shorthand of the universe. Hence, it represents a universe beyond the space we live in physically. More specifically, it is a computer-generated environment that simulates the world, which distinguishes it from metaphysical concepts\cite{Dionisio20133D}. The Concept of the Metaverse, originally from the fiction novel Snow Crash, written by Neal Stephenson in 1992\cite{Joshua2017Information}, is conveyed as a virtual world, with human-as-avatars interacting with intelligent agents and each other in that space. A more modern portrayal is discussed in the world of Ready Player One, by Ernest Cline, where users can use any role or play as any character representation in a completely virtual world\cite{Ai2021Metaverse}. Metaverse encompasses numerous definitions, and can be considered as an evolution alongside the internet.% since it is a continuing exploration term in the state-of-art. 

\begin{figure}[tb]
 \centering % avoid the use of \begin{center}...\end{center} and use \centering instead (more compact)
 \includegraphics[width=\columnwidth]{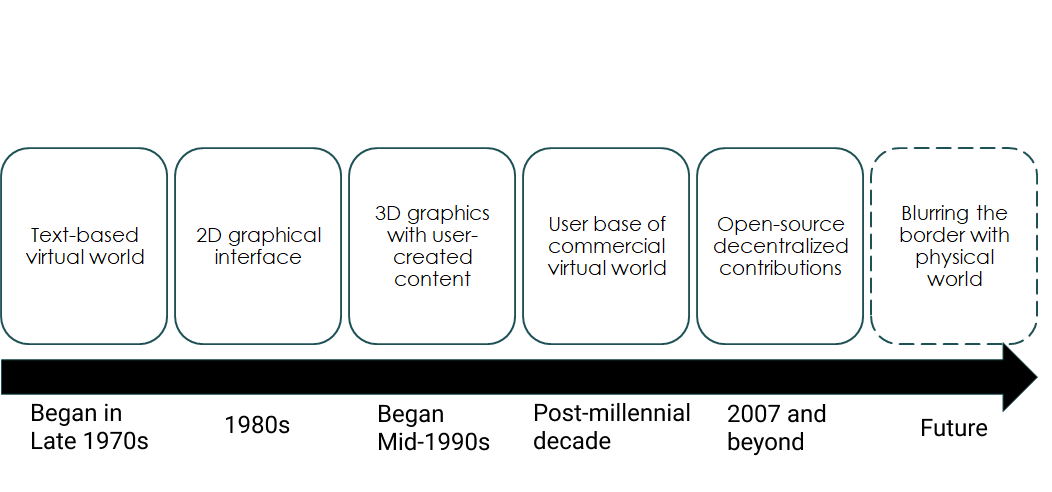}
 \caption{The Evolution of Metaverse, inspired by \cite{Dionisio20133D}}
 \label{TheEvolutionofMetaverse}
\end{figure}

The five stages of the Metaverse evolution as identified by Dionisio et al.\cite{Dionisio20133D} (see \autoref{TheEvolutionofMetaverse}) demonstrates the increasing immersive capabilities of world-building and functionality, paralleling the technological advances in multi-user systems starting from text-based computing in the late 1970s to open-source, decentralized, co-creative platforms today. The next stage of the Metaverse may be toward a close connection with many real-life applications, producing a blurring of the border with the physical world.

Similarly, Mazurek \& Gervautz\cite{Mazuryk1996Virtual} highlight that there are more terms to address within these VR themes, including Synthetic Experience, Virtual Worlds, Artificial Worlds and Artificial Reality. They indicate that although there are various definitions for VR, these meanings are equivalent to a simulated world that users could interact with and feel immersed. This highlights further properties of Metaverse environments; others such as within the AIP Cube\cite{zeltzer1992autonomy}, a taxonomy of graphic simulation systems, wherein Autonomy, Interaction, and Presence combine to describe a single interaction space: Autonomy means how well the virtual agent reacts to the simulated environment, Interaction defines the degree of users manipulating the simulated parameters, and the Presence axis measures users' perceiving system of the available sensory input and output channels. This theory applies well to AR and MR, and hence can also be considered for the blend within the Metaverse context of physical objects and virtual environments. Further concepts related to level of agency are also relevant. In terms of the agent in Metaverse, Mixed Reality Agents (MiRAs) address an agent with virtual or physical entities embodied in a Mixed Reality environment\cite{Holz2011MiRA-mixed}. The key concepts of MiRAs are Agency, Embodiment, and Interaction Capability within Mixed Reality Environments, divided into the three-axis agency, corporeal presence, and interactive capacity. Together, these each lend toward the concept of a blended virtual and physical environment, wherein agency is present, and where humans-in-the-loop can interact with the environment from a virtual-physical -- and also agent-oriented -- perspective.

\subsection{An XR-IoT (XRI) Perspective of the Metaverse}
Augmented Reality (AR) and the Internet-of-Things (IoT) are key technologies receiving significant attention, as they improve the communication between the Metaverse and physical space. AR is the interactive medium that provides computer augmented elements to the view of the real world, while IoT refers to the networking of physical objects with computing devices for sensing and communication\cite{Jo2019AR}. In \cite{Tsang2021Hybrid} this concept is described as XR-IoT (XRI), representing the combination of XR-based IoT systems as well as IoT-based XR systems. A more comprehensive system of XRI is presented by \cite{Morris2021XRI} as those that are based on immersive, information-rich, multi-user, and agent-driven environments. The potential usability scenarios of IoT and AR have been implemented in both industrial applications and academic research. The combination of these technologies make it possible to design for improving the relationships between humans and objects, human-to-human relationships, and toward future applications for a variety of domains such as education, cyber security, and marketing\cite{Andrade2019Extended}.

Similarly, XR-IoT research on IoT Avatars was presented by the authors in a simple proof-of-concept that embodies the MR representation Avatar for a physical plant, providing mixed reality interaction buttons to control physical servo motors and LEDs in the physical environment through IoT on the mobile phone\cite{Shao2019IoT}. This has been extended to explore the mixed reality framework for IoT with more interaction within an immersive Head-Mounted Display\cite{Guan2020Exploring}. The work applied stereoscopic cameras (ZED mini) attached to an HMD (Oculus Rift) to provide a video-passthrough mixed-reality experiment and collect the real-time data of the plant's context allowing for it to adapt and respond based on the following factors: intensity of  lighting, soil moisture, and the number of people in its presence. These factors informed the emotional states of a virtual plant avatar, generating a hybrid virtual-physical object with agency behaviour using fuzzy logic\cite{Morris2020Toward}. Other projects include "Digi-log", a seamless and scalable AR service and experiment for IoT-ready products, such as data visualization based on object position, the mechanism for accessing, controlling, and interacting with objects, and content exchanging interoperability, that could apply to an augmented reality shopping scenario\cite{Jo2019IoT}. A further example, HoloFlows research provided a path toward direct monitoring and control of IoT applications, %(which is limited due to the interconnectedness of today's smart devices, and HoloFlows 
introducing a new Mixed Reality interaction method for end-users to manage standard IoT devices applying a visual development interface based on Node-Red \cite{Seiger2021HoloFlows}. These examples collectively demonstrate how the IoT and XR can integrate and exchange information and foster adaptive behaviors, interfaces, and immersive visualizations.

\section{Toward an Extended Metaverse for Hyper-connected Smart environments}
\begin{figure}[tb]
 \centering % avoid the use of \begin{center}...\end{center} and use \centering instead (more compact)
 \includegraphics[width=\columnwidth]{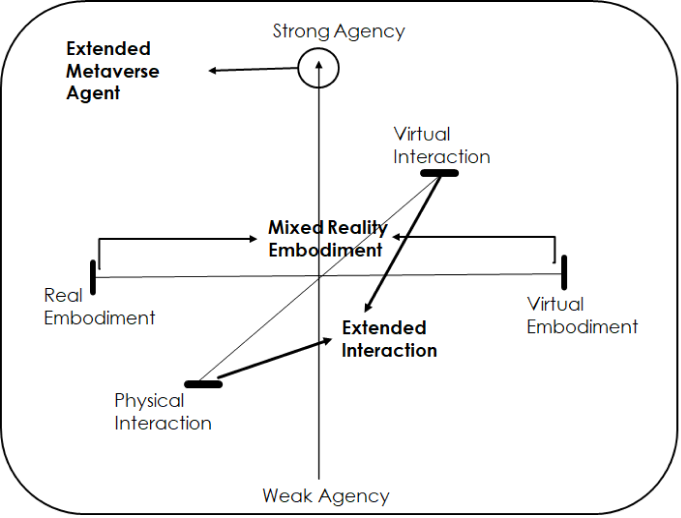}
 \caption{Criteria of Metaverse Agent (derived from the MiRAs taxonomy of \cite{Holz2011MiRA-mixed}).}
 \label{CriteriametaverseAgent}
\end{figure}

Based on a combination of theories including the MiRAs cube of \cite{Holz2011MiRA-mixed}, and the reality-virtuality continuum of Milgram \cite{Milgram1994taxonomy}, the domain of an Extended Metaverse Agent is depicted (see \autoref{CriteriametaverseAgent}) extending along the dimensions of Mixed Reality Embodiment, Extended Interaction, and level of Agency. They represent how the virtual and physical entities can be presented on the virtual or physical space, and how they could interact with users and by other agent-objects. In this sense, the Extended Metaverse consists of one or more embodied virtual and physical objects, each having a degree of interactive properties in the virtual and physical dimension and having agent-oriented behavioral capabilities. The Extended Metaverse agent facilitates the cohesive connection -- or connected-ness -- of these agents with their real-world counterparts.

\begin{figure*}[tb]
 \centering % avoid the use of \begin{center}...\end{center} and use \centering instead (more compact)
 \includegraphics[width=4.5in]{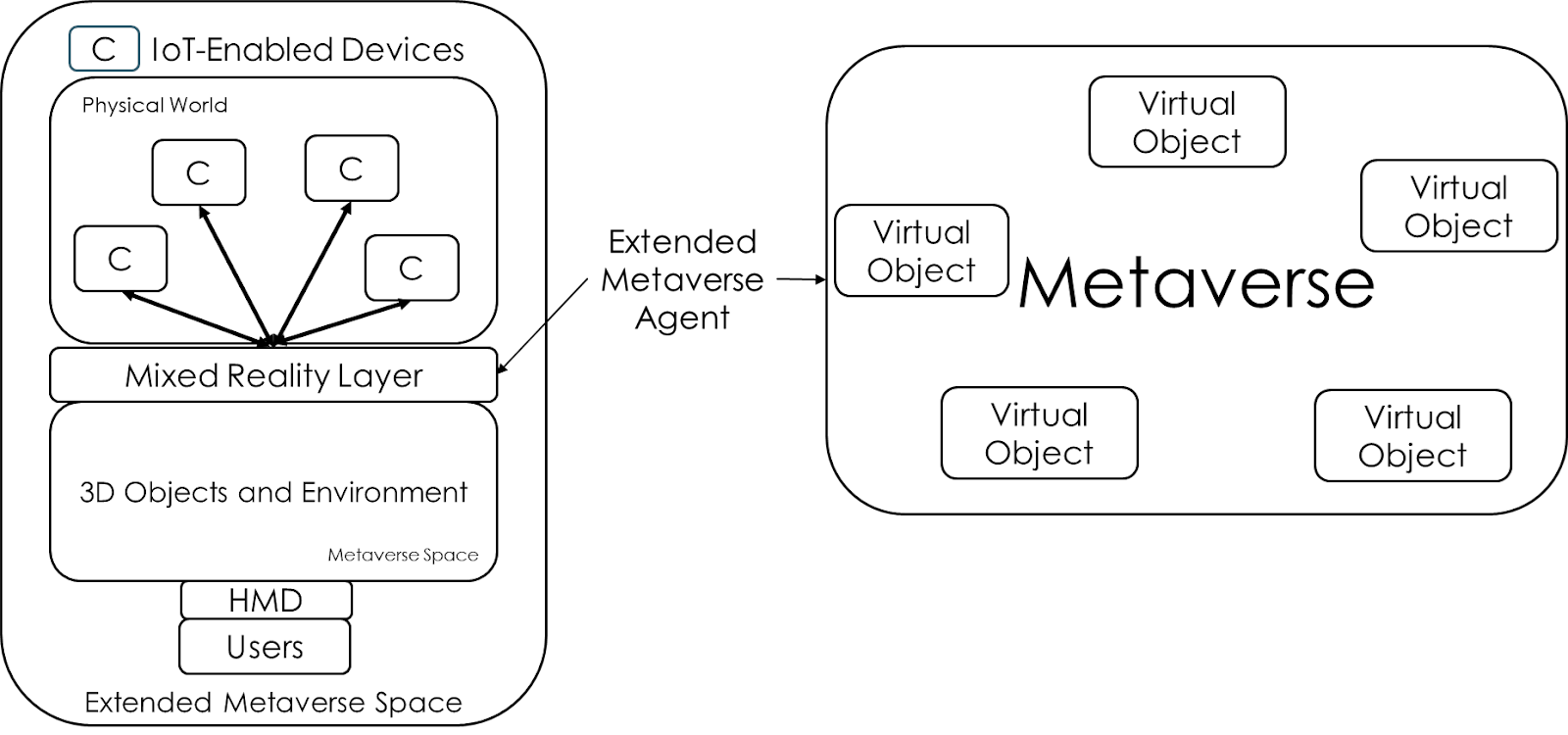}
 \caption{Framework of Extended Metaverse Agent, integrating the Metaverse of virtual objects with XR-IoT environments through an agent controller (or set of controllers), inspired by \cite{Rekimoto1995world}.}
 \label{FrameworkofExtendedMetaverseAgent}
\end{figure*}

\autoref{FrameworkofExtendedMetaverseAgent} shows the landscape for designing and creating Extended Metaverse Agents. On the physical world area, C represents the IoT-enabled devices (embedded computers), while the Extended Metaverse Agent is located on the Mixed Reality layer and connects to the virtual object in the Metaverse with a head-mounted display for users.  

\subsection{Design Prototypes Toward Extending the Metaverse}

Based on the Extended Metaverse framework described, there are multiple possible implementations, however, as an early proof-of-concept, the following are described as XRI design explorations for a more hyper-connected Metaverse, namely: an XRI Lamp controller; and an XRI ambient lighting scene. %and a work-in-progress emotional-context virtual-effect 3D scene controller.

% These scenarios provide an approach to 

\subsubsection{XRI Lamp Controller}
\begin{figure*}[tbh]
 \centering % avoid the use of \begin{center}...\end{center} and use \centering instead (more compact)
 \includegraphics[width=4.5in]{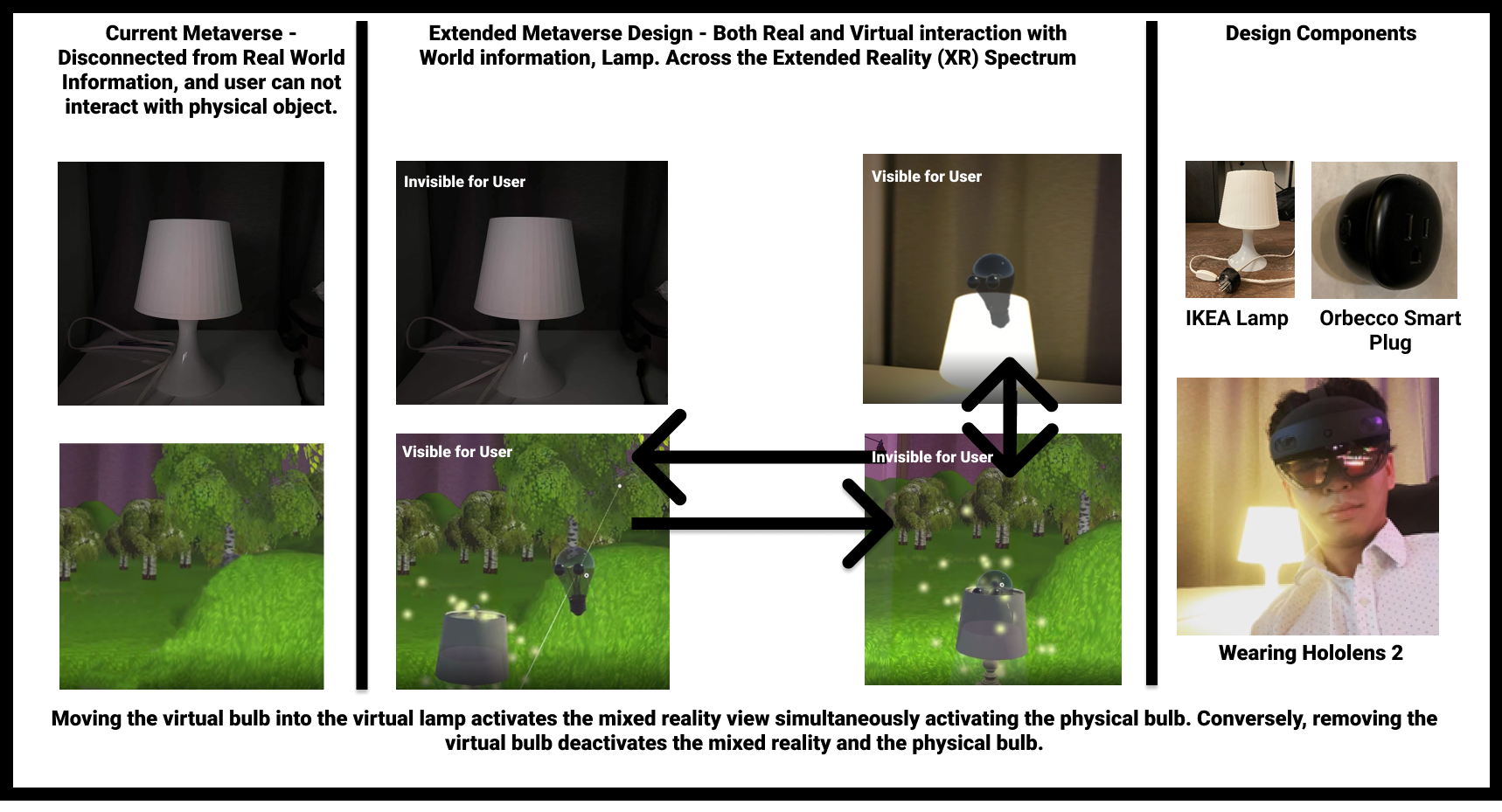}
 \caption{Design of an XRI Lamp Controller - In the extended metaverse, virtual representations adapt physical behaviors (lamp and bulb controller interface when fully immersed).
Similar mixed reality interactions take place, for controlling the virtual when in physical environment.}
 \label{SharedSmartLamp}
\end{figure*}

The XRI Lamp Controller (see \autoref{SharedSmartLamp}) is a prototype to explore a new way to control the physical object with virtual elements in Extended Reality and the possibility to switch between Virtual Reality (Metaverse Environment) and Mixed Reality space. The smart lamp is considered a shared object in Mixed Reality since it embodies both virtual and physical environments at the same position, presenting both virtual and physical properties that users are able to access. To activate the lamp, the familiar on- and off- switch nomenclature of a physical bulb is expanded to encompass either a physical or virtual action. In the virtual parameters, as the user moves a virtual bulb (virtual body presented in Virtual Reality and Mixed Reality) in and out of the physical lamp in view, the virtual interaction causes a synchronous action in both physical and virtual environments to turn the light on. The same is true in the physical parameters: as the bulb is controlled by pressing a button, the lamp is synchronized in virtual environments. In other words, the lamp object is a constant variable in both virtual and physical space, wherein the on and off function of the physical light also corresponds to control the light function in  the virtual environment. Both virtual and real environments approach a two-way dual interface functionality. 

%Note we may need to make this diagram more clear...perhaps use the video to take some screenshots and label what is happening here.

The project is constructed with the physical object, the simulated virtual entities and environment, the interaction method, head-mounted display (HMD) device, and data connection protocol. On the physical space, a consumer-grade IKEA lamp is transformed into a smart and IoT-enabled object with an Orbecco smart plug\cite{OrbeccoSmartPlug} that can connect to the Smart Life App system. The computer-simulated three-dimensional models of objects and the environment are rendered by Unity3D and visualized with Hololens 2. Meanwhile, Hololens 2 also performs hand-tracking to capture the gesture from users. The Mixed Reality Tool Kit (MRTK)\cite{MRTKMixedRealityToolkit} can use these commands to enable the interaction of the virtual elements, presented as moving the virtual bulb in this prototype. The IFTTT\cite{IFTTT} was considered as the connection service for data communication in the project since it provides a direct way to connect the Smart Life system and Webhook, which can be commanded from Unity with HTTP Request.

The prototype provides two-way data compatibility from virtual to physical commands, activating the lighting by moving the virtual bulb in and out of the constant lamp object. This  method provides a novel way to reconsider design functions, and whether it is necessary to use buttons to turn on and off the smart objects in the Extended Metaverse environment. It also  encourages designers, engineers, and users to explore how the relation with objects and virtual elements can be expanded in a new interactive dynamic ontology. The switching between Virtual Reality and Mixed Reality environments provides a bridge for users to see the physical space. However, as the current Virtual Reality headset system immerses the user into the completely computer-generated environment, it impairs the sense of vision for users to access and navigate the physical context. %As humans living in the physical realm, we cannot separate from it to enter the Metaverse Completely. Some people may argue that they can move off the VR headset from their eyes to come back to the physical area, but this project considers immanent future scenarios, where the smart glasses (or VR headset) become a part of human life like the mobile phone we use today and even a part of the human body.

\subsubsection{XRI Ambient Lighting}
\begin{figure*}[tbh]
 \centering % avoid the use of \begin{center}...\end{center} and use \centering instead (more compact)
 \includegraphics[width=5in]{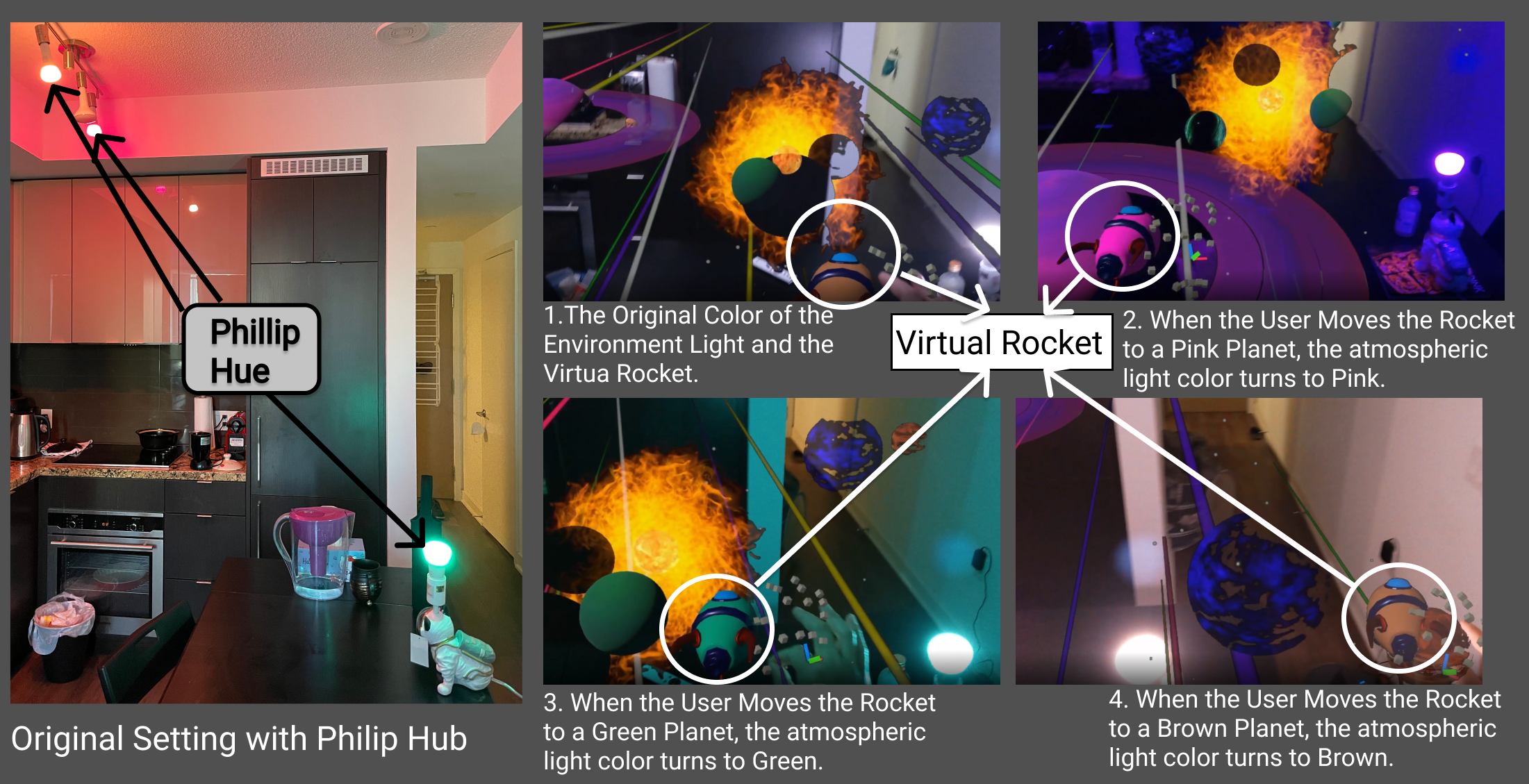}
 \caption{Design of an XRI Ambient Lighting Capability}
 \label{ColorPicker }
\end{figure*}

Another Prototype (see \autoref{ColorPicker }) explores a more dynamic way to control the physical light from Mixed Reality by changing the colour instead of only turning it on and off, namely the Mixed Reality Color Picker. When wearing the Mixed Reality headset (Hololens 2), the users are able to see some virtual planets with different colours and shapes moving around a virtual sun. The goal of the galaxy setting is to immerse the users into a Mixed Reality universe space with their everyday surroundings to transform the living room into a more dynamic and engaging space. On the physical side, four smart bulbs with colour-changing features are installed that can provide an ambience colour effect to the environment. In terms of interaction, the users can move a virtual rocket around space by air tapping and holding it with hand-tracking in the Hololens 2. If the rocket collides with the planets, it will change its colour into the colour of the planet, simultaneously changing the ambient lighting in the physical environment accordingly with the smart bulbs.

The Mixed Reality environment visualization and interaction are also made with Unity and MRTK in this prototype. The rotation of the planets (with various directions) around the sun is simply attached to a Rotator script onto the sun entity, making it the parent object of other planets. The colours of the planets are defined by their textures, while the rocket is considered as a colour picker to select the RGB value of colour; when it is moved the rocket enters into the planet with an "OnTriggerEnter" function. In terms of the physical environment lighting, it is created with the Phillips Hue White and Color Ambiance Kit\cite{philips-hue} and their Developer Kit\cite{developforHue}. %The first step to make the Phillip Hue in developer mode is to get the local IP address of the Phillip hub, which is the bridge of the bulbs to connect the local network. With the HPPT GET method, it can require the bulbs' ID and its states (On or Off, Saturation, Brightness, Hue, etc.) with JSON format. The HTTP PUT method (call from Unity on this prototype)  is for controlling the states of the bulbs, and this project focuses on changing the colour. The colour of Phillip Hue is quantified by CIE colour space\footnote{http://hyperphysics.phy-astr.gsu.edu/hbase/vision/cie.html} with X and Y values, so it needs to convert the RGB values getting from Unity to XY of CIE. The whole process of conversion is from RGB to XYZ on CIE, and then XYZ to XY of the CIE colour space. A Unity plugin (pb\_ColorUti\footnote{https://github.com/karl-/pb\_ColorUtil}) is provided to process the RGB to XYZ easily, and the calculation from XYZ to XY is processed with the formula  and \cite{Kerr2010CIE} that implemented in Unity.

Current Mixed Reality applications do not provide the capacity for users to manipulate the physical agent with the Mixed Reality Interface, and users still need to access and control the physical environment in the traditional way – physical touch. Meanwhile, although the exploration of the smart home system provides the users to monitor and control their physical IoT-enabled devices in their mobile phone APP, Shao et al. \cite{Shao2019IoT} indicates that the current IoT dashboard design is widget-based with a 2d graphic User interface and faces scaling challenges to address a large number of devices. Hence, a Mixed Reality Interface design of connect-ness to virtual elements and the physical space is identified as a critical aspect of the design process to produce dynamic, engaging, and cohesive interactive experiences. The physical ambient lighting effect provides an enhanced immersive Mixed Reality environment while combining the virtual galaxy in this prototype, as the dynamic interaction of the virtual planets and the rocket colour picker synchronously ties physical lighting colour with the virtual texture content. %In addition, the ambient lighting can enhance the immersion of the Mixed Reality environment since it changes the atmosphere of the users' surrounding that links to the virtual environment.
As this prototype presents, the ambient colour of the physical environment will change based on the planets' interaction with the rocket in the virtual realm, prompting users to connect the visual atmospheric properties as a "landing" cue on the selected planet. The user's surrounding physical atmosphere adapts to the appearance of that planet. In other words, coherence and continuity between physical and virtual spaces are hyper-connected with an extended immersive environment produced by the synchronous shared content.

The preceding design prototypes demonstrate but one approach to connect the user experience in the virtual environment of the Metaverse in a tandem relationship to their physical environment, regardless of whether this ranges from full virtual reality, to mixed reality, to non-VR applications.

\section{Discussion and Future Work}
This research has highlighted a gap in between the integration of real and virtual dimensions within the Metaverse. It proposes the following concern of a disconnect in information between the virtual and physical environments: if the Metaverse and the corresponding physical space/s are not communicating in interoperable coherence (e.g. constant objects in the physical world and the Metaverse are not compatible; or full VR immersion impairs and blocks out navigation in physical space), the ``noise'' of incongruous signals between them would potentially become overwhelming or irrelevant, perhaps eventually resulting in forms of cognitive overload for the user, from both physical and virtual domains at the same time. This situation may be a mere nuisance to the user, but as the Metaverse applications increase in prevalence and utility, such a disconnect in non-compatibility may impact important or critical application scenarios as well. %in the psychological state shutting down the overstimulation of senses of the virtual world.

% The potential for incoherence of the current Metaverse system would increase the ``noise'' of signals between physical and virtual based on the Shannon \& Weaver\cite{Shannon1964Theory} communication model, which may result in an unpredictable and uncontrollable virtual world. 

``Noise'' as defined by Shannon \& Weaver \cite{Shannon1964Theory} is specific to the engineering problem of data as signals, however the definition is relevant to today’s engineering challenges of the Metaverse. These challenges include incongruous agents that limit coherent interoperability, as well as excessive inputs in environments that limit clarity of the communications between them [e.g. cues from shared environments are inconsistent]. 
``Noise reduction'' could support the theory of reducing the gap between the Metaverse and the physical world through the design of interoperable interfaces. Designing for noise reduction would potentially have multiple benefits which span across the variety of agent interaction relationships below:

\begin{itemize}
    \item Human-to-Human Metaverse Relationships: increased efficiency with communications and tasks without the demands of attention divided in the physical and virtual world. Shared interfaces within human-to-human environment contexts foster and encourage co-creative engagement and community.
    
    \item Environment-to-Human Metaverse Relationships: as smart home and smart work environments develop with integrated IoT technology, physical environments can adapt and change to fit the needs of user or community profiles through AI and generative approaches of machine learning. Filtered data from external sensors (e.g. weather and environmental systems, transportation, social news feeds) would become naturalized elements of the explicit and implicit habitable environment.
    
    \item Object-agent to object-agent Metaverse Relationships: Dialogue between two or more objects may interact to inform users and participants of shared Metaverse spaces of collective activity or sentiment; e.g. XRI Ambient lighting conveys participant engagement, collective sentiment of community. Similarly, non-human agent-objects having presence and agency via sensors would afford communication of unseen states e.g. plants \cite{Shao2019IoT}, gaseous detection or mold, pets, etc.
\end{itemize}

As the research and development of such interfaces between agents and humans are explored, the integration of appropriate filters and controls would support the continuity and end goals of the Extended Metaverse. An Extended Metaverse Agent as proposed would help to prevent such disconnects, reducing cognitively demanding uses of Metaverse applications by improving the interaction, embodiment, and coherence of multiple Metaverse spaces, while designing for the human as a part of a dynamic system seamlessly connected across Mixed Reality Environments.

%On the one hand, the potential consequence of overwhelming "noise" of signals for the human to access as the "medium" of virtual and physical is shutting down the over-stimulation of senses in the psychological state. On the other hand, the current Metaverse is constructed by computer graphics which requires physical energy such as electric power. Hence, the overloading of information in Metaverse without rules and limitations from the physical world may result in the shutting down of Metaverse and even affect the physical life. As a consequence, if Metaverse does not have a better solution to connect and keep coherency of the physical world to provide legitimacy to the virtual simulation and evolution, the Metaverse will not be reliable for human activities.

%\subsection{Future Work}
%Note, I think we may need to more clearly relate these elements below as future work.

In terms future work, through the development of new test scenarios, the authors plan to continue these explorations to consider how the integration of the Metaverse with the actual environment can be achieved. This offers multiple promising directions toward the following:
\begin{itemize}
    \item Toward a more contextually driven usage of the user's physical and psychological contexts in the representation of the metaverse content and behavior. %Jie to integrate some of the text from the previous brain-muse application example which was removed above
    \item Toward a procedural and generative design approach for representing metaverse informational objects (this may apply generative tools like L-system representations, or others to evolve the virtual representation of the metaverse object). %Jie to integrate some of the text below.
    
    \item Toward embedding richer levels of agency into the extended metaverse designs to enable the metaverse environment's adaptive system behaviors, more autonomously across physical and virtual domains.
\end{itemize}

\section{Summary}
% What they can do?
% Address the connection problem
% Design
This work has presented a design perspective on the Metaverse, highlighting the essential need to create approaches that synchronize and connect the Metaverse more concretely to the physical world to streamline user interaction. The theoretical foundations of the Metaverse, XR, and IoT are identified, and a new Extended Metaverse framework is discussed as a means toward facilitating this strong connection. Two early-stage design projects are presented, showing how stronger connections can be made in this domain, leading to interactive and embodied Metaverse applications that adapt in both the physical and virtual realities. It is hoped that this contribution will help further the design considerations within the Metaverse research and development community as the technology moves toward mainstream acceptance and use.

\acknowledgments{
This work was supported in part by the Tri-council of Canada under the Canada Research Chairs program.}

\bibliographystyle{abbrv-doi}

\bibliography{template}
\end{document}